\documentclass[a4paper]{jpconf}
\usepackage{amsmath,bm,amssymb}
\usepackage[square,sort&compress]{natbib}

\usepackage{graphicx}

\newcommand{\bk}{b_{\bm{k}}}

\newcommand{\bkd}{b_{\bm{k}}^{\dag}}

\newcommand{\bmk}{b_{-\bm{k}}}

\newcommand{\ckl}{c_{\bm{k},\lambda}}

\newcommand{\ckld}{c_{\bm{k},\lambda}^{\dag}}

\newcommand{\sumk}{\sum_{\bm{k}}}
\newcommand{\sumkl}{\sum_{\bm{k},\lambda}}

\newcommand{\wk}{w_{\bm{k}}}

\newcommand{\omk}{\omega_{\bm{k}}}

\newcommand{\fk}{f_{\bm{k}}}

\newcommand{\gkl}{g_{\bm{k}\lambda}}

\newcommand{\kk}{\bm{k}}

\newcommand{\rl}{\rangle\!\langle}

\begin{document}

\title{Radiative and phonon-induced dephasing in double quantum dots}

\author{Pawe{\l} Machnikowski}

\address{Institute of Physics, Wroc{\l}aw University of Technology,
  50-370 Wroc{\l}aw, Poland}

\ead{Pawel.Machnikowski@pwr.wroc.pl}

\begin{abstract}
A simple method for describing the evolution of a quantum state of a double quantum dot
system interacting simultaneously with the electromagnetic environment
and with the lattice modes is developed. It is shown that the joint
action of the two reservoirs leads to nontrivial effects in the system
dephasing. As an example, the impact of phonon-induced initial
dephasing on the radiative decay of delocalized exciton states is discussed.
\end{abstract}

\section{Introduction}

Double quantum dots (DQDs) are systems composed of two quantum dots
placed close enough to each other for new coherent and collective
phenomena to occur which cannot be reduced to the known properties of
individual dots. Such new phenomena related to the optical
properties of the system include modification of the optical
response due to inter-dot coupling \cite{unold05,danckwerts06} or 
collective effects in the spontaneous emission
\cite{sitek07a,sitek09a}. Another class of effects is related with
carrier-phonon interaction. Examples include phonon-assisted exciton
transfer between the dots \cite{govorov03,richter06,rozbicki08a}, modification
of optical spectra due to phonon packets traveling between the dots
\cite{huneke08}, or 
phonon-induced decay of entanglement \cite{roszak06a}. The special
properties of DQDs may open the way to new applications, like
long-time storage of 
quantum information \cite{pazy01b}, conditional optical
control \cite{unold05} that may lead to an implementation of a two-qubit quantum
gate \cite{biolatti00}, generation of entangled photons
\cite{gywat02} or coherent optical spin control and entangling
\cite{troiani03,nazir04,gauger08b}. Although both spontaneous emission
and phonon-related phenomena have been studied to some extent for DQD
systems, the interplay between these two classes of effects has not
been investigated.

In the present contribution, we formulate a theoretical description
of the evolution of a double quantum dot coupled simultaneously to the
two reservoirs: the quantum electromagnetic field (photon vacuum) and lattice
vibrational modes (phonons). After defining the model
(Sec.~\ref{sec:system}), the approach to the simulation of the
quantum open system is presented (Sec.~\ref{sec:evol}), followed by
an example of an interplay between the two couplings
(Sec.~\ref{sec:interplay}).

\section{The system}
\label{sec:system}

The system under study is composed of two stacked self-assembled QDs
with transition energies 
$\epsilon_{1}$ and $\epsilon_{2}$, 
interacting with their phonon and photon (radiative)
environments. We restrict the discussion to the ground states of
excitons in each dot and assume that the spin polarizations of the
excitons are fixed. As the exciton dissociation energy in absence of
external electric fields is rather large (several to a few tens of
meV), we consider only spatially direct exciton states, i.e., such
that the electron-hole pairs reside in one and the same dot.

Thus, the model includes four basis states: the state without
excitons, $|0\rangle$, the states $|1\rangle$ and $|2\rangle$ with an
exciton in the first and second dot, respectively, and the `molecular
biexciton' state $|3\rangle$ with excitons in both dots.
The system evolution will be described in a `rotating basis' defined by the
unitary transformation 
\begin{equation*}
U=
e^{iE t(|1\rl 1|+|2\rl 2|+2|3\rl 3|)/\hbar},
\end{equation*}
where $E=(\epsilon_{1}+\epsilon_{2})/2$. Note that this is not
equivalent to interaction picture with respect to the Hamiltonian of
uncoupled dots. Upon the standard weak-coupling derivation of the
Master equation, the latter would always lead to uncorrelated emission
for non-identical dots. On the contrary, the present transformation
allows us to explicitly keep the energy difference 
$\Delta=(\epsilon_{1}-\epsilon_{2})/2$ which
yields unitary evolution superposed on the dissipative one,
consistent with the result of the Weisskopf--Wigner approach
\cite{sitek07a}, and allows one to correctly account for the transition from
``identical dots'' to ``different dots''. 
The Hamiltonian is then
\begin{equation*}
H=H_{\mathrm{DQD}}
+H_{\mathrm{ph}}+H_{\mathrm{rad}}
+H_{\mathrm{c-ph}}+H_{\mathrm{c-rad}}.
\end{equation*}

The first term describes exciton states in the DQD structure,
\begin{equation}\label{ham-DQD}
H_{\mathrm{DQD}} =  \Delta(|1\rl 1|-|2\rl 2|)
+V(|1\rl 2|+|2\rl 1|),
\end{equation}
where $V$ is the coupling between the
dots, which may originate either from the Coulomb (F{\"o}rster)
interaction or from tunnel coupling (it can be assumed real). 
It is convenient to introduce the parametrization
$\Delta=\mathcal{E}\cos 2\theta,\quad 
V=\mathcal{E}\sin 2\theta$,
where $2\mathcal{E}$ is the energy splitting between the single-exciton
eigenstates of $H_{\mathrm{DQD}}$ and $\theta$ is the mixing
angle of the single-exciton states.  
Electron and hole wave functions are modelled by identical
anisotropic Gaussians 
with identical extensions $l$ in the $xy$ plane
and $l_{z}$ along $z$ for both particles,
\begin{displaymath}
\psi_{1,2}(\bm{r})\sim\exp\left[ 
-\frac{1}{2}\frac{x^{2}+y^{2}}{l^{2}} - \frac{1}{2}\frac{(z\pm D/2)^{2}}{l_{z}^{2}}
 \right],
\end{displaymath}
where $D$ is the distance between the dots.

The phonon modes are described by the free phonon Hamiltonian
$H_{\mathrm{ph}}=\sumk\hbar\omk\bkd\bk$,
where $\bk,\bkd$ are bosonic
operators of the phonon modes and $\omk$ are the corresponding frequencies.
Interaction of carriers confined in the
DQD with phonons is modelled by the independent boson Hamiltonian
\begin{equation}\label{ham-phon}
H_{\mathrm{c-ph}}=(|1\rl 1|+|3\rl|3|)\sumk\fk^{(1)}(\bkd+\bmk)
+(|2\rl 2|+|3\rl|3|)\sumk\fk^{(2)}(\bkd+\bmk), 
\end{equation}
where $\fk^{(1,2)}$
are system-reservoir coupling constants.
For Gaussian wave functions, the coupling constants for the
deformation potential coupling between 
confined charges and longitudinal phonon modes have the form 
$\fk^{(1,2)}=\fk e^{\pm ik_{z}D/2}$, where 
\begin{displaymath}
\fk=(\sigma_{\mathrm{e}}-\sigma_{\mathrm{h}})\sqrt{\frac{k}{2\varrho vc_{\mathrm{l}}}}
\exp\left[
-\frac{l_{z}^{2}k_{z}^{2}+l^{2}k_{\bot}^{2}}{4}\right].
\end{displaymath}
Here $v$ is the normalization volume,  
$k_{\bot/z}$ are momentum
components in the $xy$ plane and along the $z$ axis,
$\sigma_{\mathrm{e/h}}$ are deformation potential constants for
electrons/holes, $c_{\mathrm{l}}$ is the speed of longitudinal sound,
and $\varrho$ is the crystal density. We assume that off-diagonal
carrier--phonon couplings are negligible due to small overlap of the
wave functions confined in different dots.

The third component in our modeling is the radiative reservoir (modes
of the electromagnetic field), described by the Hamiltonian
$H_{\mathrm{rad}}=\sumkl\hbar\wk\ckld\ckl$,
where $\ckl,\ckld$ are photon creation and annihilation operators
and $\wk$ are the corresponding frequencies ($\lambda$ denotes
polarizations). 
The QDs
are separated by a distance much smaller than the relevant photon
wavelength $\lambda=2\pi\hbar c/E$, so that the spatial dependence of
the EM field may be neglected (the Dicke limit). 
The Hamiltonian describing the
interaction of carriers with the 
EM modes in the dipole and rotating wave approximations is 
\begin{equation}\label{hI}
H_{\mathrm{c-rad}}  = 
\Sigma_{-}\sumkl e^{-iEt/\hbar}\gkl \ckld +\mathrm{H.c.},
\end{equation}
with $\Sigma_{-}
=|0\rl 1|+|2\rl 3| + |0\rl 2| + |1\rl 3|$
and
$\gkl=i\bm{d}\cdot\hat{e}_{\lambda}(\kk)
\sqrt{\frac{\hbar\wk}{2\varepsilon_{0}\varepsilon_{\mathrm{r}}v}}$,
where 
$\bm{d}$ is the interband dipole moment, $\varepsilon_{0}$ is the
vacuum permittivity, $\varepsilon_{\mathrm{r}}$ is the dielectric
constant of the semiconductor and
$\hat{e}_{\lambda}(\kk)$ is the unit polarization vector
of the photon mode with the wave vector $\kk$ and polarization
$\lambda$.  
For wide-gap semiconductors with $E\sim 1$ eV, zero-temperature
approximation may be used for the radiation reservoir at any
reasonable temperature. 

\section{The system evolution}
\label{sec:evol}

In certain limiting cases, analytical formulas for the evolution of
the DQD system may be found. For uncoupled dots ($V=0$) interacting
only with lattice modes (phonons), an exact solution is available
\cite{roszak06a}. If only the radiative decay is included, a solution
in the Markov limit can be obtained \cite{sitek07a}. Here, we propose
a general description which allows one to deal with the simultaneous action
of both these environments. We describe the evolution of the reduced
density matrix of the DQD system in the interaction
picture with respect to $H_{\mathrm{DQD}}$ by the equation
\begin{equation*}
\dot{\rho}=\mathcal{L}_{\mathrm{rad}}[\rho]
+\mathcal{L}_{\mathrm{ph}}[\rho].
\end{equation*}
Here the first term describes the effect of the radiative decoherence
in the Markovian limit in terms of the Lindblad dissipator
\begin{equation*}
\mathcal{L}_{\mathrm{rad}}[\rho]=\Gamma_{\mathrm{rad}} 
\left[ \Sigma_{-}(t)\rho\Sigma_{+}(t)
-\frac{1}{2}\{\Sigma_{+}(t)\Sigma_{-}(t),\rho \}_{+}\right],
\end{equation*}
where $\Sigma_{-}(t)=\Sigma_{+}^{\dag}(t)
=e^{iH_{\mathrm{DQD}}t/\hbar}\Sigma_{-}e^{-iH_{\mathrm{DQD}}t/\hbar}$
and 
$\Gamma_{\mathrm{rad}}=E^{3}|\bm{d}|^{2}\sqrt{\epsilon_{\mathrm{r}}}/
(3\pi\epsilon_{0}c^{3}\hbar^{4})$
is the spontaneous decay rate for a single dot. 
The second term accounts for the interaction with the non-Markovian
phonon reservoir. We use the time-convolutionless equation
\begin{equation}\label{tcl}
\mathcal{L}_{\mathrm{ph}}[\rho]=
-\int_{0}^{t}d\tau\tr_{\mathrm{ph}}\left[ 
H_{\mathrm{c-ph}}(t),\left[ H_{\mathrm{c-ph}}(\tau),\rho(t)\otimes\rho_{\mathrm{ph}} 
\right]  \right],
\end{equation}
where $H_{\mathrm{c-ph}}(t)
=e^{i(H_{\mathrm{DQD}}+H_{\mathrm{ph}})t/\hbar}
H_{\mathrm{c-ph}}e^{-i(H_{\mathrm{DQD}}+H_{\mathrm{ph}})t/\hbar}$
is the carrier-phonon interaction Hamiltonian in the interaction picture,
$\rho_{\mathrm{ph}}$ is the phonon density matrix at the thermal
equilibrium, and $\tr_{\mathrm{ph}}$ denotes partial trace with
respect to phonon degrees of freedom. While some phonon--assisted
processes allow a Markovian limit, there are many effects that can
only be reproduced if the reservoir memory is included (such as the
initial phonon--assisted dephasing). Moreover, there seems to be no
universal way to extract the Markov limit in various physical
situations. By using the general Eq.~(\ref{tcl}), we obtain an
all-purpose equation of motion at a modest computational cost. 

In the evolution generated by the $H_{\mathrm{DQD}}$, the states
$|0\rangle$ and $|3\rangle$ are invariant and nontrivial evolution
takes place only in the subspace spanned by $|1\rangle,|2\rangle$. The
corresponding evolution operator can therefore be found easily.
Transforming Eq.~(\ref{ham-phon}) to the interaction picture and
substituting to Eq.~(\ref{tcl}) we find the phonon-related
contribution to the system evolution in the explicit form
\begin{displaymath}
\mathcal{L}_{\mathrm{ph}}[\rho]=
\sum_{i=1,2}\left[ T_{i}(t)\rho(t)S_{i}(t)-S_{i}(t)T_{i}(t)\rho(t)
\right] +\mathrm{H.c.}
\end{displaymath}
Here the nonzero elements of the matrices $S_{i}$ are
\begin{eqnarray*}
\langle 1|S_{1}|1\rangle=1-\langle 2|S_{1}|2\rangle
=1-\langle 1|S_{2}|1\rangle=\langle 2|S_{2}|2\rangle
& = & \frac{1}{2}\sin^{2}2\theta\left(
  1-\cos\frac{2\mathcal{E}t}{\hbar}
 \right), \\
\langle 1|S_{1}|2\rangle=\langle 2|S_{1}|1\rangle^{*}
=-\langle 1|S_{2}|2\rangle=-\langle 2|S_{2}|1\rangle^{*}
& = &\frac{i}{2}\sin 2\theta\sin\frac{2\mathcal{E}t}{\hbar}
-\frac{1}{4}\sin 4\theta\left( 1-\cos\frac{2\mathcal{E}t}{\hbar} \right),\\
\langle 3|S_{1}|3\rangle=\langle 3|S_{2}|3\rangle & = & 1,
\end{eqnarray*}
and the matrices $T_{i}(t)$ are defined by
\begin{displaymath}
T_{i}(t)=\sum_{j=1,2}\int_{0}^{t}d\tau S_{j}(\tau)R_{ij}(t-\tau).
\end{displaymath}
Here 
\begin{eqnarray*}
R_{11}(t)=R_{22}(t) & = & \frac{1}{\hbar^{2}}\sumk|\fk|^{2}
\left[ n_{\bm{k}}e^{i\omk t}+(n_{\bm{k}}+1)e^{-i\omk t} \right],\\
R_{12}(t)=R_{21}(t) & = & \frac{1}{\hbar^{2}}\sumk|\fk|^{2}\cos k_{z}D
\left[ n_{\bm{k}}e^{i\omk t}+(n_{\bm{k}}+1)e^{-i\omk t} \right]
\end{eqnarray*}
are memory functions with $n_{\kk}$ denoting the Bose distribution for
phonon modes.
The above equation of motion for the reduced density matrix strictly
reproduces the results in the limiting cases mentioned
above. Moreover, for the case of a DQD coupled to phonons with
non-vanishing inter-dot coupling $V$, it yields results reasonably close to
those obtained by a much more complex correlation expansion technique
\cite{rozbicki08a}. 

In numerical simulations, we take the parameters corresponding to a self-assembled
InAs/GaAs system: $\sigma_{\mathrm{e}}-\sigma_{\mathrm{h}}=9$~eV,
$\rho=5350$~kg/m$^{3}$, $c_{\mathrm{l}}=5150$~m/s, 
the wave function parameters
$l=4.5$~nm, $l_{z}=1$~nm, $D=6$~nm, and the radiative recombination time (for a
single dot) $1/\Gamma_{\mathrm{rad}}=400$~ps.

\section{Example: Impact of phonon-induced pure dephasing on
  spontaneous emission} 
\label{sec:interplay}

\begin{figure}[tb]
\centering
\includegraphics[width=120mm]{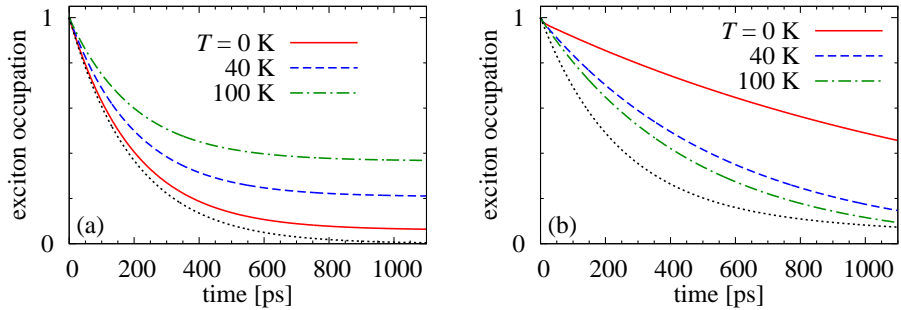}\hspace{2pc}%
\caption{\label{label}Decay of the exciton
    occupation for the superradiant initial state $|\psi_{+}\rangle$
    in the absence of phonon-induced dephasing (black dotted lines) and in the
    presence of phonon-induced dephasing at various temperatures as
    shown. (a) identical, uncoupled dots, $\Delta=V=0$; (b)
    non-identical, coupled dots, $\Delta=V=1$~meV.} 
\end{figure}

As an example of an application of the formalism developed above, let
us study a simple example of interplay between phonon-related and
radiative phenomena. First, we consider identical, uncoupled dots, that is,
$V=0$, $\Delta=0$. In such case of two identical emitters, the
spontaneous emission has a collective character which leads to a strong
modification of the exciton recombination process \cite{sitek07a}. In
particular, out of the two delocalized single exciton states
$|\psi_{\pm}\rangle=(|1\rangle\pm|2\rangle)/\sqrt{2}$,
$|\psi_{+}\rangle$ has a ``superradiant'' character and 
decays twice faster than a single-dot excitation, while $|\psi_{-}\rangle$ is
stable against recombination (``subradiant''). In Fig.~\ref{label}(a),
the decay of the state $|\psi_{+}\rangle$ in the absence of phonon
perturbation is shown by the black dotted line. 

For a single dot, phonon-induced dephasing (within an independent
boson model with a harmonic reservoir) leads only to
coherence decay on picosecond time scales and does not affect the
radiative recombination on long time scales. This situation changes in
quite an interesting manner in the two-dot case. As one can see in
Fig.~\ref{label}(a), in the presence of phonon-induced dephasing, the
exciton recombination slows down and a long-lived occupation appears
in the system. Moreover, the final exciton occupation grows as the
temperature increases. Obviously, this effect must be due to some kind
of coherence leading to occupation trapping in a subradiant state
which is only possible in such a clear form because the model does not
contain dephasing channels that would destroy this coherence (like
anharmonicity-induced scattering \cite{machnikowski06d}). Still, such
a noise-induced coherence is remarkable.

The reason for this behavior is that during the first few picoseconds
of the open system evolution the coherence between the single exciton
states decays due to carrier-phonon dynamics according to
$\langle 1|\rho|2\rangle = \frac{1}{2}e^{-h(t)}$,
where $\rho$ is the density matrix of the exciton subsystem and 
\begin{displaymath}
h(t)=4\sumk\left|\frac{\fk}{\omk}\right|^{2} \sin^{2}\frac{k_{z}D}{2}
(1-\cos\omk t)\stackrel{t\gg t_{\mathrm{ph}}}{\longrightarrow}
h_{\infty}=4\sumk\left|\frac{\fk}{\omk}\right|^{2} \sin^{2}\frac{k_{z}D}{2},
\end{displaymath}
where $t_{\mathrm{ph}}$ is the characteristic time of the
phonon-induced initial dephasing which is of the order of a few
picoseconds \cite{roszak06a}. Actually, the above formulas are valid
in the absence of 
spontaneous emission. However, on the time scales $t\ll\Gamma$, when
the dephasing takes place, radiative contribution can be approximately
neglected. 
As a result, the initial density matrix $\rho=|\psi_{+}\rl\psi_{+}|$ is
transformed into
\begin{displaymath}
\rho'=\frac{1+e^{-h_{\infty}}}{2}|\psi_{+}\rl\psi_{+}|
+\frac{1-e^{-h_{\infty}}}{2}|\psi_{-}\rl\psi_{-}|.
\end{displaymath}
This dephased state is a mixture of a superradiant and a subradiant
component. The latter does not decay radiatively and leads to the
persistent occupation tail visible in Fig.~\ref{label}(a).

In the more realistic case of dots that differ in their transition
energies and are coupled [Fig.~\ref{label}(b)], the radiative
recombination process still depends on temperature but this dependence
is reversed: Now, the emission speeds up as the temperature is
increased. In any case, however, the decay in the presence of
carrier-phonon coupling is slower than in the absence of phonons. In
order to explain this behavior, we note that although the eigenstates of
$H_{\mathrm{DQD}}$ in the presence of energy mismatch are not purely
sub- and superradiant one of them still decays faster than the other
\cite{sitek07a}. For $V>0$, the lower eigenstate has a partly
subradiant character. Moreover, the coupling between the dots
enables excitation transfer between these eigenstates
\cite{rozbicki08a}. Coupling to phonons provides a thermalization
mechanism for the occupations of the two eigenstates. At $T=0$ this
means a transfer to the lower-energy, subradiant state, which strongly
suppresses emission. At higher temperatures, however, the contribution
from the higher-energy, superradiant state increases and the radiative
decay speeds up.

\section{Conclusion}

Double quantum dots not only show a richer structure of exciton
states than a single dot but
also admit much more complicated dephasing channels. The approach to
numerical simulation of the quantum open system dynamics of a single
DQD structure presented in this contribution opens the way to a
systematic study of these effects. Although much more accurate
(sometimes even exact) methods can be used in the case of two dots
interacting with a single reservoir
\cite{krugel06,richter06,rozbicki08a} treating the evolution under
joint action of both reservoirs at the same level is either impossible
or very demanding numerically. The method presented here, based on the
TCL equation for the non-Markovian phonon-related effects and the
Lindblad equation for spontaneous emission, yields simple equations
for the complete biexciton density matrix of the DQD system and allows
one to calculate arbitrary exciton-related properties of an undriven
system (or a system excited with a single ultrashort pulse).

\ack
This work was supported by the Polish MNiSW under Grant No. N N202 1336
33. 


\providecommand{\newblock}{}


\begin{thebibliography}{10}
\expandafter\ifx\csname url\endcsname\relax
  \def\url#1{{\tt #1}}\fi
\expandafter\ifx\csname urlprefix\endcsname\relax\def\urlprefix{URL }\fi
\providecommand{\eprint}[2][]{\url{#2}}

\bibitem{unold05}
Unold T, Mueller K, Lienau C, Elsaesser T and Wieck A~D 2005 {\em Phys. Rev.
  Lett.\/} {\bf 94} 137404

\bibitem{danckwerts06}
Danckwerts J, Ahn K~J, F{\"o}rstner J and Knorr A 2006 {\em Phys. Rev. B\/}
  {\bf 73} 165318

\bibitem{sitek07a}
Sitek A and Machnikowski P 2007 {\em Phys. Rev. B\/} {\bf 75} 035328

\bibitem{sitek09a}
Sitek A and Machnikowski P 2009 {\em Phys. Rev. B\/} {\bf 80} 115301

\bibitem{rozbicki08a}
Rozbicki E and Machnikowski P 2008 {\em Phys. Rev. Lett.\/} {\bf 100} 027401

\bibitem{govorov03}
Govorov A~O 2003 {\em Phys. Rev. B\/} {\bf 68} 075315

\bibitem{richter06}
Richter M, Ahn K~J, Knorr A, Schliwa A, Bimberg D, Madjet M~E~A and Renger T
  2006 {\em Phys. Stat. Sol. (b)\/} {\bf 243} 2302

\bibitem{huneke08}
Huneke J, Kr\"{u}gel A, Kuhn T, Vagov A and Axt V~M 2008 {\em Phys. Rev. B\/}
  {\bf 78} 085316

\bibitem{roszak06a}
Roszak K and Machnikowski P 2006 {\em Phys. Rev. A\/} {\bf 73} 022313

\bibitem{pazy01b}
Pazy E, D'Amico I, Zanardi P and Rossi F 2001 {\em Phys. Rev. B\/} {\bf 64}
  195320

\bibitem{biolatti00}
Biolatti E, Iotti R~C, Zanardi P and Rossi F 2000 {\em Phys. Rev. Lett.\/} {\bf
  85} 5647

\bibitem{gywat02}
Gywat O, Burkard G and Loss D 2002 {\em Phys. Rev. B\/} {\bf 65} 205329

\bibitem{troiani03}
Troiani F, Molinari E and Hohenester U 2003 {\em Phys. Rev. Lett.\/} {\bf 90}
  206802

\bibitem{nazir04}
Nazir A, Lovett B~W, Barrett S~D, Spiller T~P and Briggs G~A~D 2004 {\em Phys.
  Rev. Lett.\/} {\bf 93} 150502

\bibitem{gauger08b}
Gauger E~M, Nazir A, Benjamin S~C, Stace T~M and Lovett B~W 2008 {\em New J.
  Phys.\/} {\bf 10} 073016

\bibitem{machnikowski06d}
Machnikowski P 2006 {\em Phys. Rev. Lett.\/} {\bf 96} 140405

\bibitem{krugel06}
Kr{\"u}gel A, Axt V~M and Kuhn T 2006 {\em Phys. Rev. B\/} {\bf 73} 035302

\end{thebibliography}
\end{document}